\documentclass[journal,11pt, onecolumn]{IEEEtran}
\usepackage{microtype}

\usepackage{times,epsfig,nicefrac,color,balance}
\usepackage{graphics, mathtools, cuted}

\usepackage{cite,calc}

\usepackage{amsthm}
\usepackage{amsmath}
\usepackage{amssymb}
\usepackage{mathtools}
\usepackage{mathrsfs}

\usepackage{multirow}
\usepackage{array}

\usepackage{amsfonts}
\usepackage{graphicx}
\usepackage[linesnumbered]{algorithm2e}

\usepackage{color, verbatim}
\usepackage[usenames,dvipsnames]{xcolor}
\usepackage{url}
\usepackage{cleveref}

\usepackage{epsfig}
\usepackage{subfig}

\newcommand\nc\newcommand
\nc{\bb}[1]{\mathbb{#1}}
\renewcommand{\cal}[1]{\mathcal{#1}}
\renewcommand{\bf}[1]{\mathbf{#1}}

\DeclarePairedDelimiter{\set}{\lbrace}{\rbrace}
\DeclarePairedDelimiter{\br}{\lparen}{\rparen}
\DeclarePairedDelimiter{\brac}{\lbrack}{\rbrack}
\DeclarePairedDelimiter{\abs}{\lvert}{\rvert}
\nc\defeq{\mathrel{\mathop:}=}

\DeclareMathOperator{\spn}{span}

\nc\bfa{{\boldsymbol a}}\nc\bfA{{\boldsymbol A}}\nc\cA{{\cal A}} \nc\fA[1]{A\br*{#1}} \nc\fa[1]{a\br*{#1}}  \nc\rmA{\mathrm{A}} \nc\rma{\mathrm{a}}
\nc\bfb{{\boldsymbol b}}\nc\bfB{{\boldsymbol B}}\nc\cB{{\cal B}} \nc\fB[1]{B\br*{#1}} \nc\fb[1]{b\br*{#1}}  \nc\rmB{\mathrm{B}} \nc\rmb{\mathrm{b}}
\nc\bfc{{\boldsymbol c}}\nc\bfC{{\boldsymbol C}}\nc\cC{{\cal C}} \nc\fC[1]{C\br*{#1}} \nc\fc[1]{c\br*{#1}}  \nc\rmC{\mathrm{C}} \nc\rmc{\mathrm{c}}
\nc\bfd{{\boldsymbol d}}\nc\bfD{{\boldsymbol D}}\nc\cD{{\cal D}} \nc\fD[1]{D\br*{#1}} \nc\fd[1]{d\br*{#1}}  \nc\rmD{\mathrm{D}} \nc\rmd{\mathrm{d}}
\nc\bfe{{\boldsymbol e}}\nc\bfE{{\boldsymbol E}}\nc\cE{{\cal E}} \nc\fE[1]{E\br*{#1}} \nc\fe[1]{e\br*{#1}}  \nc\rmE{\mathrm{E}} \nc\rme{\mathrm{e}}
\nc\bff{{\boldsymbol f}}\nc\bfF{{\boldsymbol F}}\nc\cF{{\cal F}} \nc\fF[1]{F\br*{#1}} \nc\ff[1]{f\br*{#1}}  \nc\rmF{\mathrm{F}} \nc\rmf{\mathrm{f}}
\nc\bfg{{\boldsymbol g}}\nc\bfG{{\boldsymbol G}}\nc\cG{{\cal G}} \nc\fG[1]{G\br*{#1}} \nc\fg[1]{g\br*{#1}}  \nc\rmG{\mathrm{G}} \nc\rmg{\mathrm{g}}
\nc\bfh{{\boldsymbol h}}\nc\bfH{{\boldsymbol H}}\nc\cH{{\cal H}} \nc\fH[1]{H\br*{#1}} \nc\fh[1]{h\br*{#1}}  \nc\rmH{\mathrm{H}} \nc\rmh{\mathrm{h}}
\nc\bfi{{\boldsymbol i}}\nc\bfI{{\boldsymbol I}}\nc\cI{{\cal I}} \nc\fI[1]{I\br*{#1}} \nc\rmI{\mathrm{I}} \nc\rmi{\mathrm{i}}
\nc\bfj{{\boldsymbol j}}\nc\bfJ{{\boldsymbol J}}\nc\cJ{{\cal J}} \nc\fJ[1]{J\br*{#1}} \nc\fj[1]{j\br*{#1}} \nc\rmJ{\mathrm{J}} \nc\rmj{\mathrm{j}}
\nc\bfk{{\boldsymbol k}}\nc\bfK{{\boldsymbol K}}\nc\cK{{\cal K}} \nc\fK[1]{K\br*{#1}} \nc\fk[1]{k\br*{#1}} \nc\rmK{\mathrm{K}} \nc\rmk{\mathrm{k}}
\nc\bfl{{\boldsymbol l}}\nc\bfL{{\boldsymbol L}}\nc\cL{{\cal L}} \nc\fL[1]{L\br*{#1}} \nc\fl[1]{l\br*{#1}} \nc\rmL{\mathrm{L}} \nc\rml{\mathrm{l}}
\nc\bfm{{\boldsymbol m}}\nc\bfM{{\boldsymbol M}}\nc\cM{{\cal M}} \nc\fM[1]{M\br*{#1}} \nc\fm[1]{m\br*{#1}} \nc\rmM{\mathrm{M}} \nc\rmm{\mathrm{m}}
\nc\bfn{{\boldsymbol n}}\nc\bfN{{\boldsymbol N}}\nc\cN{{\cal N}} \nc\fN[1]{N\br*{#1}} \nc\fn[1]{n\br*{#1}} \nc\rmN{\mathrm{N}} \nc\rmn{\mathrm{n}}
\nc\bfo{{\boldsymbol o}}\nc\bfO{{\boldsymbol O}}\nc\cO{{\cal O}} \nc\fO[1]{O\br*{#1}} \nc\fo[1]{o\br*{#1}} \nc\rmO{\mathrm{O}} \nc\rmo{\mathrm{o}}
\nc\bfp{{\boldsymbol p}}\nc\bfP{{\boldsymbol P}}\nc\cP{{\cal P}} \nc\fP[1]{P\br*{#1}} \nc\fp[1]{p\br*{#1}} \nc\rmP{\mathrm{P}} \nc\rmp{\mathrm{p}}
\nc\bfq{{\boldsymbol q}}\nc\bfQ{{\boldsymbol Q}}\nc\cQ{{\cal Q}} \nc\fQ[1]{Q\br*{#1}} \nc\fq[1]{q\br*{#1}} \nc\rmQ{\mathrm{Q}} \nc\rmq{\mathrm{q}}
\nc\bfr{{\boldsymbol r}}\nc\bfR{{\boldsymbol R}}\nc\cR{{\cal R}} \nc\fR[1]{R\br*{#1}} \nc\fr[1]{r\br*{#1}} \nc\rmR{\mathrm{R}} \nc\rmr{\mathrm{r}}
\nc\bfs{{\boldsymbol s}}\nc\bfS{{\boldsymbol S}}\nc\cS{{\cal S}} \nc\fS[1]{S\br*{#1}} \nc\fs[1]{s\br*{#1}} \nc\rmS{\mathrm{S}} \nc\rms{\mathrm{s}}
\nc\bft{{\boldsymbol t}}\nc\bfT{{\boldsymbol T}}\nc\cT{{\cal T}} \nc\fT[1]{T\br*{#1}} \nc\ft[1]{t\br*{#1}} \nc\rmT{\mathrm{T}} \nc\rmt{\mathrm{t}}
\nc\bfu{{\boldsymbol u}}\nc\bfU{{\boldsymbol U}}\nc\cU{{\cal U}} \nc\fU[1]{U\br*{#1}} \nc\fu[1]{u\br*{#1}} \nc\rmU{\mathrm{U}} \nc\rmu{\mathrm{u}}
\nc\bfv{{\boldsymbol v}}\nc\bfV{{\boldsymbol V}}\nc\cV{{\cal V}} \nc\fV[1]{V\br*{#1}} \nc\fv[1]{v\br*{#1}} \nc\rmV{\mathrm{V}} \nc\rmv{\mathrm{v}}
\nc\bfw{{\boldsymbol w}}\nc\bfW{{\boldsymbol W}}\nc\cW{{\cal W}} \nc\fW[1]{W\br*{#1}} \nc\fw[1]{w\br*{#1}} \nc\rmW{\mathrm{W}} \nc\rmw{\mathrm{w}}
\nc\bfx{{\boldsymbol x}}\nc\bfX{{\boldsymbol X}}\nc\cX{{\cal X}} \nc\fX[1]{X\br*{#1}} \nc\fx[1]{x\br*{#1}} \nc\rmX{\mathrm{X}} \nc\rmx{\mathrm{x}}
\nc\bfy{{\boldsymbol y}}\nc\bfY{{\boldsymbol Y}}\nc\cY{{\cal Y}} \nc\fY[1]{Y\br*{#1}} \nc\fy[1]{y\br*{#1}} \nc\rmY{\mathrm{Y}} \nc\rmy{\mathrm{y}}
\nc\bfz{{\boldsymbol z}}\nc\bfZ{{\boldsymbol Z}}\nc\cZ{{\cal Z}} \nc\fZ[1]{Z\br*{#1}} \nc\fz[1]{z\br*{#1}} \nc\rmZ{\mathrm{Z}} \nc\rmz{\mathrm{z}}

\nc\flog[1]{\log\br*{#1}}
\nc\fexp[1]{\exp\br*{#1}}

\newcommand\F{{\mathbb F}}

\newtheorem{theorem}{Theorem}
 \newtheorem{definition}{Definition}
\crefname{definition}{definition}{definitions}
\newtheorem{lemma}[theorem]{Lemma}

\newcommand{\remove}[1]{}

\crefname{Appendix}{Appendix}{Appendices}

\begin{document}
\sloppy
\title{Local Partial Clique Covers for Index Coding}
\author{Abhishek Agarwal  and Arya Mazumdar
	\thanks{Abhishek Agarwal is with the Department of Electrical and Computer Engineering, University of Minnesota, Minneapolis, MN  55455. Arya Mazumdar is with the Computer Science Department of University of Massachusetts, Amherst, MA 01003, and was with the University of Minnesota. email: \texttt{abhiag@umn.edu, arya@cs.umass.edu}. Research supported by NSF grants CCF 1318093 and CAREER CCF 1453121.}
}

%
%
%
\allowdisplaybreaks
\maketitle

\begin{abstract}
Index coding, or broadcasting with side information, is a network coding problem of most fundamental importance. 
In this problem, given a directed graph, each vertex represents a user with a need of information, and the neighborhood of each vertex  represents the side information availability to that  user. The  aim is to find an encoding  to minimum number of bits (optimal rate)
that, when broadcasted, will be sufficient to the need of every user. Not only the optimal rate is intractable, but it is also very hard to characterize with some other well-studied graph parameter or with a simpler formulation, such as a linear 
program. Recently there have been a series of works that address this question and provide explicit schemes for index coding as the optimal value of a linear program with rate
given by  well-studied
properties such as local chromatic number or partial clique-covering number. 
There has been a recent attempt to combine these existing  notions of local chromatic number  and partial clique covering into a unified notion denoted as the {\em local partial clique cover} (Arbabjolfaei and Kim, 2014).

We present a generalized novel upper-bound (encoding scheme) - in the form of the minimum value of a linear program - for optimal index coding. Our bound also combines the    notions of local chromatic number  and partial clique covering into a new definition of the  local partial clique cover, which outperforms both the previous bounds, as well as beats the previous attempt to combination.

  Further, we look at the upper bound derived recently by Thapa et al., 2015,  and extend their $n$-$\mathsf{GIC}$ (Generalized Interlinked Cycle) construction to $(k,n)$-$\mathsf{GIC}$ graphs, which are a generalization of $k$-partial cliques.  
\end{abstract}


\section{Introduction}

Index coding is a multiuser communication problem in which the broadcaster reduces the total number of transmissions by taking advantage of the fact that a user may have side information about other user's data. We consider the index coding problem formulation of \cite{bar2011index}. 
Suppose, a set of users are assigned bijectively
to a set of vectors that they want to know. However, instead
of knowing the assigned vector, each knows a subset of
other vectors. This scenario can be depicted by a so-called
side-information graph where each vertex represents a user
and there is an edge from user A to user B, if A knows
the vector assigned to B. Given this graph, how much
information should a broadcaster transmit, such that
each vertex can deduce its assigned vector?

\begin{figure}
  \centering
 \includegraphics[width=6cm]{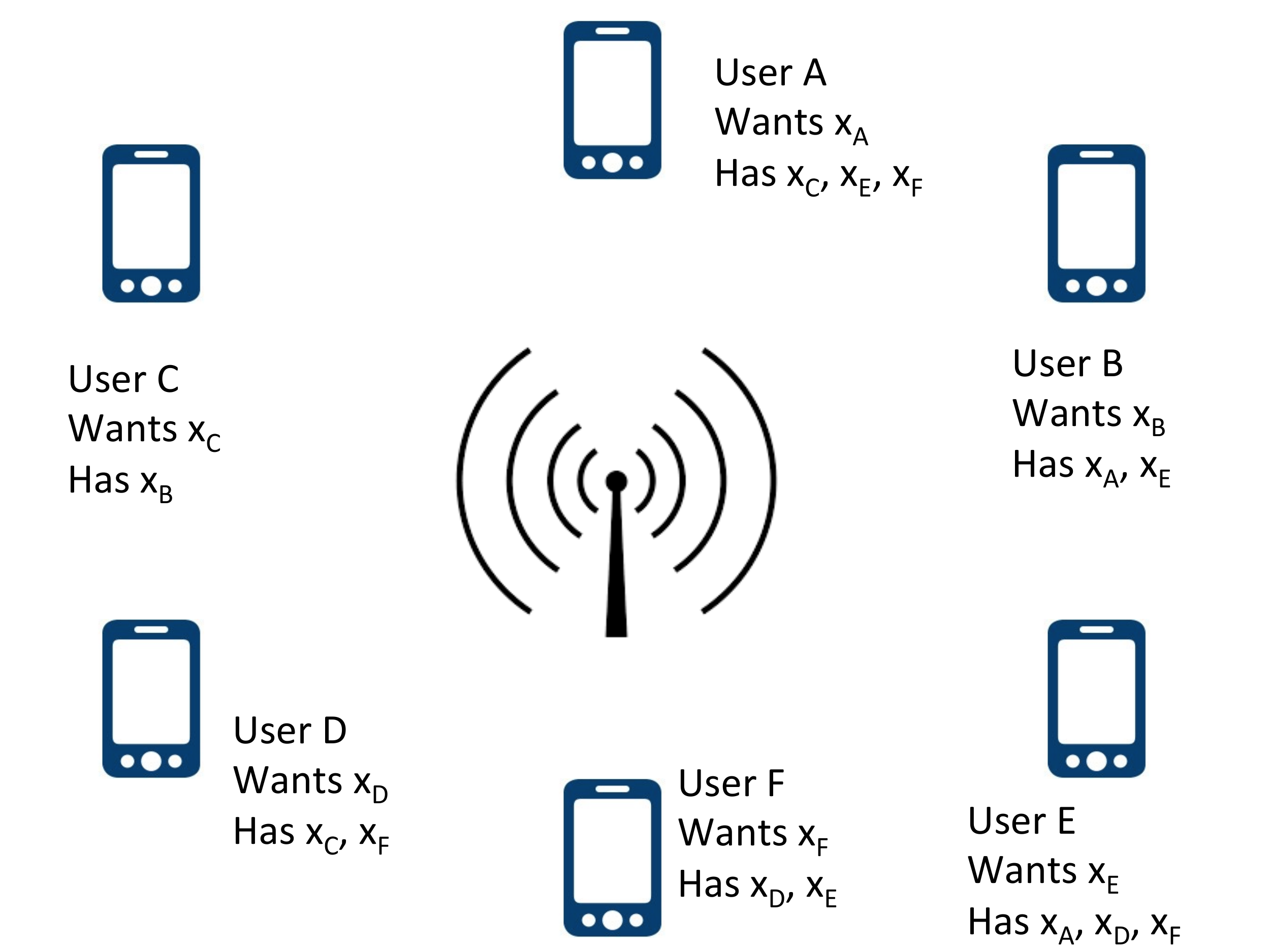}
\caption{Example of an Index Coding problem. The corresponding side-information graph is depicted in \cref{fig:local_example}}	
\label{fig:index_example}
\end{figure}

\begin{figure}
  \centering
 \includegraphics[width=6cm]{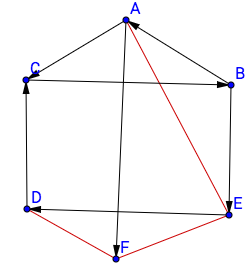}
\caption{Example of an Index Coding problem where the fractional local chromatic number and the fractional partial clique number based index codes are strictly larger than the proposed scheme for the fractional case. The undirected edges denote  bidirectional edges.}	
\label{fig:local_example}
\end{figure}

Consider a group of $n$ users with the side information represented by the directed graph $\cG$, such that the out-neighbors of vertex $v_i \in V(\cG) = \set{v_1, v_2, \ldots, v_n}$ denote the nodes whose data is available to $v_i$. The data requested by node $v_i$ is represented by a vector $\bf{x}_i \in \F_q^\ell$. Let $N(v,\cG) \subseteq V(\cG)\setminus \set{v}$ denote the out-neighbors of vertex $v$ ($v\in \overline{N(v,\cG)}$). For example, an index coding scenario has been depicted in \cref{fig:index_example}. The side information graph for this scenario is given in  \cref{fig:local_example}.
In the graph $\cG$ of  \cref{fig:local_example}, $N(A, \cG) = \{C,E,F\}, N(B, \cG) = \{A,E\}, N(C, \cG) = \{B\}, N(D, \cG) = \{C,F\}, N(E, \cG)= \{A,D,F\}, N(F, \cG)= \{D, E\}$.
The aim is to broadcast the minimum amount of data (in-terms of $\F_q^\ell$-symbols) such that vertex $v_i$ is able to reconstruct $\bf{x}_i$ from ${\bf{x}}_{N(v_i,\cG)}$ (vectors assigned to the neighbors) and the broadcast for all $i \in \{1, \dots, n\}.$ The amount of broadcasted data is referred to as the {\em broadcast rate} of the index coding problem.

This simple formulation attracted substantial attention recently \cite{alon2008broadcasting,lubetzky2009nonlinear,blasiak2011lexicographic}. In particular, it has been shown that any network coding problem can be cast as an  index coding problem \cite{el2010index,effros2015equivalence}. Furthermore, index coding has been identified as the hardest instance of all of
network coding \cite{langberg2008hardness}. For up to $n=5$, the optimal broadcast rates of all index coding problems have been tabulated in \cite{YHIndexWeb}. 

If the recovery functions for the nodes, as well as the encoding function at the broadcaster are all linear over $\F_q$, then the optimal (minimum)
broadcasting rate is given by a quantity called the {\em minrank} of the graph \cite{haemers1978, bar2011index}. In general, minrank is  hard to compute for arbitrary graphs, and also does not give an achievable scheme in terms of simpler well known graph parameters.

There have been a series of works that aim to characterize the index coding broadcast rate in terms of other possibly intractable graph parameters or cast the index coding rate
as the optimal solution  to a linear program.  To start with, an immediate achievable scheme in terms of the {\em clique cover} number the graph or the {\em chromatic number} of the complementary graph was provided in \cite{birk1998informed}. Their simple idea is to partition the graph into minimum number of vertex-disjoint cliques. And then for each 
clique, the broadcaster transmits only once, that is the sum (as $\F_q$-vectors) of the data  in all the vertices of the clique. This provides
an index coding solution equal to the chromatic number of the complementary graph. 

The above approach then was extended in  \cite{shanmugam2013local}. In particular,  in \cite{shanmugam2013local} 
an interference alignment perspective of the linear index coding problem was given.
It was  shown that, it is possible to achieve an index coding solution with number of transmissions equal to the
{\em local chromatic number}  \cite{erdos1986coloring} of the graph.
There also exists an orthogonal approach to generalize the clique covering bound (or the chromatic number bound). 
In this approach, appearing as early as in \cite{birk1998informed}, one finds a {\em partial clique} cover (covering with subgraphs instead of cliques)
and uses maximum-distance separable (MDS) codes for each of the partial cliques.

More recently, in \cite{arbabjolfaei2014local}, the above two approaches were merged and a combined linear program solution was proposed that we discuss later.
In this paper, we show that these two orthogonal approaches can be merged more effectively, and provide a modified definition of {\em local partial clique cover}. This lead
to better achievability bound for linear index codes than any previous approaches. 

In addition, recently another approach to provide achievable index coding schemes is shown in \cite{thapa2015generalized}, where the graph is 
covered by  structures called {\em generalized interlinked cycles (GIC)}. We are also able to generalize this method in the way of partial clique covers.

The paper is organized as follows. In  \cref{sec:prelim} we define the necessary graph parameters, describe the previous schemes and narrate our main theorems. In this section we also give some concrete examples to show the better performance of our solution.   In \cref{sec:gic} we provide some further results by showing a generalization of the GIC scheme of \cite{thapa2015generalized}.
In \cref{sec:achiev_sch} we describe our main achievability scheme (and hence the proof of the main result) that outperforms the previous results. This is followed up in \cref{sec:frac} where we provide a linear programming (LP) relaxation. Proofs and the example have been shortened for space constraints. 

\section{Preliminaries and main result}
\label{sec:prelim}

\subsection{Notations}
The following set of notations will be useful throughout the paper, and we provide it at the outset for easy reference.
\begin{itemize}
	\item $[n] \equiv \set{1,2,\ldots,n}$
	\item $[m,n] \equiv \set{m,m+1,\ldots,n}, m\leq n$
	\item Complement of set $A$, is denoted by $\overline{A}$
	\item For a graph $\cG$, $V(\cG)$ and $E(\cG)$ denote the vertex and the edge set of the graph, respectively. $\overline{\cG}$ denotes the directed complement of $\cG$
	\item  For any set $A = \set{i_1, i_2, \ldots, i_r} \subseteq [n]$ and set of vectors $\set{\bf{v}_i}_{i\in [n]}$, $\bf{v}_{A}$denotes the set $\set{\bf{v}_{j}}_{j\in A}$ and $\bf{v}_{[A]}$ denotes the matrix $[\bf{v}_{i_1}\; \bf{v}_{i_2}\; \ldots \bf{v}_{i_r}]$. For a matrix $G \in \F^{k\times n}$, $G_{[B]}$ denotes the sub-matrix of $G$ constructed from the columns of $G$ corresponding to $B \subseteq [n]$.
	\item For a graph $\cG(V,E)$ and a set $S\subseteq V$, $\cG\rvert_S$ denotes the subgraph induced by $S$.
	\item   For a graph $\cG(V,E)$, $N(v,\cG) \subseteq V(\cG)\setminus \set{v}$ denotes the set of out-neighbors of $v\in V$.
  \item $I(\text{expr})$ denotes  the indicator function for $expr$.
  \item An $m \times n$ matrix, $m \le n$ is MDS if any $m$ columns of the matrix are linearly independent. 
  \item A linear code is called MDS code, if the generator (or the parity-check) matrix of the code is MDS. An $[n,k]$-MDS code implies the code is of length $n$ and dimension $k$.
  \item For a matrix $M$, $M(i,j)$ denotes the entry at $i^{th}$ row and $j^{th}$ column.
\end{itemize}

We will need the following definitions. 
{\definition \label{def:coloring} For a directed graph $\cG$, a coloring of indices is proper if for every vertex $v$ of $\cG$ the color any of its out-neighbors, $N(v,\cG)$, is different from the color of vertex $v$. The minimum number of colors in a proper coloring is the chromatic number $\chi$.}

{\definition \label{def:loc_coloring} The local chromatic number $\chi_l$ of a directed graph $\cG$ is the maximum number of colors in any out-neighborhood (including itself) of a vertex,
minimized over all proper colorings of the undirected graph obtained from $\cG$ by ignoring
the orientation of edges in $\cG$.}

The following definition of index coding and the optimal broadcast rate will be used throughout the paper.
  In the index coding problem, a  {\em directed} { side information} graph
$\cG(V,E)$ is given. Each vertex $v_i \in V$ represents a receiver that is interested in knowing a uniform random vector $\bfx_i \in \F^\ell_q$.
 The receiver at $v$ knows the values of the  variables $\bfx_u, u \in N(v,\cG)$.
\begin{definition}[Index coding]
An  {\em index
code} $\cC$ for $\F_q^{\ell \times n}$ with side information graph $\cG(V,E), V = \{v_1,v_2,\dots,v_n\},$ is a set of
codewords (vectors) in $\F_q^{\ell \cdot n_0}$ together with:
\begin{enumerate}
\item An encoding function $f$ mapping inputs in $\F_q^{\ell \times n}$
to codewords, and
\item A set of deterministic decoding functions $g_1,\dots,g_n$ such
that $g_i\Big(f(\bfx_1,\dots,\bfx_n), \{\bfx_j: j \in N(v_i, \cG)\}\Big) = \bfx_i$ for every $i=1, \dots,n$.
\end{enumerate}
 The encoding and decoding functions 
 depend on $\cG$. The number $n_0$ is called  the broadcast rate of the index code $\cC$ and $\ell$ is called the {\em size} of a node. Note that $n_0$ can be a non-integer as long as $n_0 \cdot \ell$ is integer. The minimum value  of $n_0$ over all possible index codes and all possible $\ell$ is called the optimal broadcast rate. 
\end{definition} 

\subsection{Prior work}

An {\em interference alignment} perspective to the graph coloring approach described in the introduction 
was presented in \cite{shanmugam2013local}
 by assigning a vector $\bf{v}_i$ to $v_i$ such that,
\begin{equation}\label{interference_align_condition}
	\bf{v}_i \not\in \spn(\bf{v}_{\overline{N(v_i,\cG)}\setminus \{v_i\}}).
\end{equation}
From the interference alignment perspective, $\overline{N(v,\cG)}\setminus\{v\}$ are the interfering set of indices for user $v$. Let $G = [\bf{v}_1 \bf{v}_2 \ldots \bf{v}_n]$. Then, the index code (the broadcaster transmission) is given by $ G\cdot [\bf{x}_1\; \bf{x}_2\; \ldots \bf{x}_n]_{\ell \times n}^T \in \F_q^{\ell \times rank(G)}$. It can be seen that each node $v_i$ can recover $\bf{x}_i$ from the index code because of \cref{interference_align_condition}. We also sometime denote the vector assigned to node $v \in V(\cG)$ as $\bf{v}(v)$.

It is possible to satisfy the requirements in \cref{interference_align_condition} if $\spn(\bf{v}_{[n]}) = n$. The goal is to minimize the dimension of $\spn(\bf{v}_{[n]})$. One solution to this problem is to find a  {{proper coloring}} (see, definition \ref{def:coloring}) of the vertices in $\overline{\cG}$ and assign orthonormal vectors to each color  (the same vector is assigned to all vertices with the same color). Thus, an achievable solution is given by the chromatic number of ${\overline{\cG}}$, $\chi(\overline{\cG})$.

One way to improve the upper-bound $\chi(\overline{\cG})$ above is to use an $[n,k]$-MDS code with $k=\chi_l(\overline{\cG})$. 

 The local chromatic number of $\overline{\cG}$ can be given by the following integer program (IP),
\begin{equation}\label{local_chrom_num_integer_prog}
\begin{array}{ll@{}}
\text{minimize}  & t\\
\text{subject to}& \sum_{\cS \in \cK_0: \cS \not\subseteq N(v,\cG)} \rho_\cS \le t, \;\; v\in V(\cG)   \\
\text{and}& \displaystyle\sum_{\cS \in \cK_0 : v \in \cS}  \rho_\cS \geq 1, \;\; v\in V(\cG) \\
                 &                                          \rho_\cS \in \{0,1\},\;\; \cS \in \cK_0
\end{array}
\end{equation}
where $\cK_0$  denotes the set of cliques in the directed graph $\cG$. Denote $\rho^{\ast}_{\cS}$ as the optimal solution to the above IP. Note that there exists a proper coloring of $\bar{\cG}$ corresponding to the IP in \cref{local_chrom_num_integer_prog}. We use the generator matrix of a $[\sum_{\cS \in \cK_0}\rho^{\ast}_{\cS},\chi_l(\overline{\cG})]$-MDS code to assign a column vector to each color in the proper coloring. In \cite{shanmugam2013local}, it was shown that this scheme works for every directed graph by assigning each vertex a vector corresponding to its color in the proper coloring of \cref{local_chrom_num_integer_prog}. 

\begin{theorem}[Shanmugam et al. \cite{shanmugam2013local}] \label{thm:shanm}The minimum broadcast rate of an index coding problem is upper bounded by the 
optimal value of the integer program of \cref{local_chrom_num_integer_prog}.
\end{theorem}

Note that, a coloring of $\overline{\cG}$ is equivalent to a clique covering of $\cG$. Thus, another  approach to finding an index code is to find a {\em partial clique}  (\cref{def:k_partial_clique}) cover of $\cG$ \cite{birk1998informed}. 
{\definition \label[definition]{def:k_partial_clique} A directed graph $\cG$ on $n$ vertices is a $k$-partial clique if every vertex has at least $n-1-k$ out-neighbors and there is at least one vertex in $\cG$ with exactly $n-k-1$ out-neighbors}.

In each of the $k_\cS$-partial clique $\cS$, one might use a parity check matrix of a Reed-Solomon code that can correct $k_\cS+1$ erasures. The broadcaster finds the syndrome of the symbols at the vertices. The dimension of the syndrome is $k_\cS+1$.
Thus, the broadcast rate of the index coding problem is given by the following integer program,
\begin{equation}\label{partial_cliq_cover_integer_prog}
\begin{array}{ll@{}}
\text{minimize}  & \displaystyle\sum_{\cS \in \cK} (k_\cS+1)\rho_\cS \\
\text{subject to}& \displaystyle\sum_{\cS \in \cK : v \in \cS}  \rho_\cS \geq 1,\;\;  v \in V(\cG)  \\
                 &                                          \rho_\cS \in \{0,1\}, \;\;\cS \in \cK
\end{array}
\end{equation}
where $\cK$ denotes the power set of vertices in $\cG$ and $\cS \in \cK$ is a $k_\cS$-partial clique.

Subsequently in this paper we combine the schemes in \cite{shanmugam2013local} and \cite{birk1998informed} to derive an achievability scheme that would give better results than both of these existing schemes. Recently, in an effort to combination, Arbabjolfaei and Kim \cite{arbabjolfaei2014local} 
proposes the following result. 
\begin{theorem}[\label{thm:arbabjolfaei2014local} Arbabjolfaei and Kim \cite{arbabjolfaei2014local}] The minimum broadcast rate of an index coding problem on the
side information graph $\cG$ is upper bounded by the optimal value of the following integer program,
\begin{equation}\label{eq:arbabjolfaei}
\begin{array}{ll@{}}
\text{minimize}  & t\\
\text{subject to}& \sum_{\cS \in \cK: \cS \not\subseteq N(v,\cG)} (k_\cS+1)\rho_\cS \le t, \;\; v\in V(\cG)   \\
\text{and}&   \displaystyle\sum_{\cS \in \cK : v \in \cS}  \rho_\cS \geq 1,\;\;  v \in V(\cG)  \\
                 &                                          \rho_\cS \in \{0,1\}, \;\;\cS \in \cK.
\end{array}
\end{equation}
\end{theorem}
We provide a stronger bound than \cref{eq:arbabjolfaei} (see \cref{thm:main} below). Our bound requires a significantly involved  achievability scheme.

There is another approach to combine the above two methods of local chromatic number and partial clique covering proposed in \cite{shanmugam2014}. 
The approach in \cite{shanmugam2014} uses the local chromatic number on partitions of the side information graph. The partitions are decided such that the sum of the local chromatic number of the complement graph of the partitions is minimized.
However
for the unicast index coding problem that we consider,  \Cref{thm:arbabjolfaei2014local} can be shown to outperform this scheme of
\cite{shanmugam2014}.

Very recently another technique for index coding achievability was proposed in \cite{thapa2015generalized}. We refrain from separately describing the scheme
here. Instead in \cref{sec:gic}, we provide a generalization of their scheme. The generalization is similar in spirit to that of $k$-partial clique covering
from clique covering. 

\subsection{Our main results}
Our main result is given by the following theorem.
\begin{theorem}\label{thm:main}
The minimum broadcast rate of an index coding problem on the
side information graph $\cG$ is upper bounded by the optimum value of the following integer program.
\begin{subequations}\label{local_k_partial_integer_prog}
\begin{flalign}
\text{minimize}  & \;\;\;t \nonumber\\
\text{subject to}& \sum_{\cS \in \cK} \min\set{\abs{\cS\cap \overline{N(v,\cG)}},(k_\cS+1)}\rho_\cS \le t, \label{local_k_partial_integer_prog_min_max_constraint}\\
 & \hspace{2in} v \in V(\cG)  \nonumber \\ 
\text{and}& \displaystyle\sum_{\cS \in \cK : v \in \cS}  \rho_\cS \geq 1, \;\; v \in V(\cG)  \label{local_k_partial_integer_prog_cover_constraint}\\
                 &                                          \rho_\cS \in \{0,1\}, \;\;\cS \in \cK.
\end{flalign}
\end{subequations}
\end{theorem}
Any vertex subgraph of a $k$-partial clique is also a $k$-partial clique. Thus, the IP in \cref{local_k_partial_integer_prog} finds a partition of $\cG$ into partial cliques i.e. the inequality in \cref{local_k_partial_integer_prog_cover_constraint} can be replaced by equality. 
We call the optimal solution to \cref{local_k_partial_integer_prog} the {\em local partial clique cover} number.

To construct a scheme to achieve the bound in \cref{thm:main} we assign vectors to each vertex $u\in V\br{\cG}$, $\bf{v}_u$, such that the condition in \cref{interference_align_condition} is satisfied. And for each partial clique we use the column vectors of a generator matrix of an $(\abs{\cS},k_\cS+1)$-MDS code. The length of the index code is equal to the span of the vectors $\set{\bf{v}_u}_{u\in V(\cG)}$. The detailed scheme is described in \cref{sec:achiev_sch}. 

The term $\min\set{\abs{\cS\cap \overline{N(v,\cG)}},(k_\cS+1)}$ in \cref{local_k_partial_integer_prog_min_max_constraint} denotes the span of the non-neighbors of vertex $v$ in the partition $\cS$. Note that, we count at most $(k_{\cal{S}}+1)$ non-neighbors in a $k_{\cal{S}}$-partial clique because the span of the vectors corresponding to $\cS$ is $k_\cS+1$.

To find the local chromatic number, we partition the graph into cliques and minimize the maximum number of partitions in the neighborhood of any vertex in the complementary side information graph, $\bar{\cG}$. In \cref{thm:main}, we replace the clique cover with a partial clique cover and minimize the maximum span of the vectors in the neighborhood of any vertex in $\overline{\cG}$. This gives a generalization of the local chromatic number of $\overline{\cG}$ for partial clique cover, and reduces to $\chi_l\br{\cG}$ when only $0$-partial cliques are used. Since a clique cover is the special case of a $k$-partial cover (a clique is a $0$-partial clique), \cref{local_k_partial_integer_prog} trivially outperforms \cref{local_chrom_num_integer_prog,partial_cliq_cover_integer_prog}. 

The  change in \cref{local_k_partial_integer_prog} from \cref{eq:arbabjolfaei} is introduction of the term $\abs{\cS\cap \overline{N(v,\cG)}}$ in the optimization under the summation. Although the set over which the summation is performed seemed to have expanded, it is not the case. Notice that, if $S \subseteq N(v, \cG)$ then $\abs{\cS\cap \overline{N(v,\cG)}} =0$. This also shows that our scheme is at least as good as \cref{eq:arbabjolfaei}. 
However the achievability scheme becomes significantly complicated because of the introduction of the term under the summation. 

In comparison, for the linear program in \cref{eq:arbabjolfaei}, a simpler achievability scheme is possible and we describe it in \cref{appendix_b} for completeness. Note that, this simple achievability scheme was not mentioned in \cite{arbabjolfaei2014local}.


We can consider the fractional relaxation of the IP in \cref{local_k_partial_integer_prog} and provide an even better upper-bound.
\begin{equation}\label{local_k_partial_linear_prog}
\begin{array}{ll@{}}
\text{minimize}  & t \\
\text{subject to}& \sum_{\cS \in \cK} \min\set{\abs{\cS\cap \overline{N(v,\cG)}},(k_\cS+1)}\rho_\cS \le t, \\
&\hspace{2in} v \in V(\cG)  \\ 
\text{and}& \displaystyle\sum_{\cS \in \cK : v \in \cS}  \rho_\cS \geq 1, \;\; v \in V(\cG)  \\
                 &                                          \rho_\cS \in [0,1], \;\;\cS \in \cK.
\end{array}
\end{equation}
\begin{theorem}[Linear programming relaxation]\label{thm:frac}
The minimum broadcast rate of an index coding problem on the
side information graph $\cG$ is at the most the optimum value of the linear programming of \cref{local_k_partial_linear_prog}.
\end{theorem}
The necessary adjustments required for the proof of this theorem is provided in \cref{sec:frac}.

Following the ideas of \cite{arbabjolfaei2014local}, we can further tighten the achievability bounds by considering recursive linear program.

\begin{theorem}[Recursive LP]\label{thm:recur}
 Let $IC_{FLP}(\cG)$ denote the optimizated value of the  linear program below for graph $\cG$:
\begin{equation}\label{local_partial_rec_lin_prog}
\begin{array}{ll@{}}
\text{minimize}  & t \\
\text{subject to}& \sum_{\cS \in \cK} \min\set{\abs{\cS\cap \overline{N(v,\cG)}},IC_{FLP}(\cG\rvert_\cS)}\rho_\cS \le t, \\
 & \hspace{2in} v \in V(\cG)  \\ 
\text{and}& \displaystyle\sum_{\cS \in \cK : v \in \cS}  \rho_\cS \geq 1, \;\; v \in V(\cG)  \\
                 &                                          \rho_\cS \in [0,1], \;\;\cS \in \cK.
\end{array}
\end{equation}
Also define $IC_{FLP}$ to be equal to $1$ for singleton sets. The minimum broadcast rate of an index coding problem on the
side information graph $\cG$ is bounded from above by $IC_{FLP}(\cG)$.
\end{theorem}
The proof of this theorem is going to be evident at the end of \cref{sec:achiev_sch} and  \cref{sec:frac}.
\vspace{0.1in}
\subsubsection{Examples of better performance for our scheme}

While it had been evident that our proposed bounds are at least as good as the previous bounds, we now give explicit example of graphs where our upper bounds are strictly
smaller than all previous upper bounds. Consider the index coding problem described by the graph in \cref{fig:local_example}. For this graph, the index coding based on the fractional local chromatic number (LP relaxation of \cref{thm:shanm}) has broadcast rate $4$,  the index coding based on the fractional partial clique clique covering (LP relaxation of \cref{partial_cliq_cover_integer_prog}) has broadcast rate $11/3$ and our proposed scheme (\cref{thm:frac}) has  broadcast rate $7/2$.

\begin{figure}
  \centering
 \includegraphics[width=9.5cm]{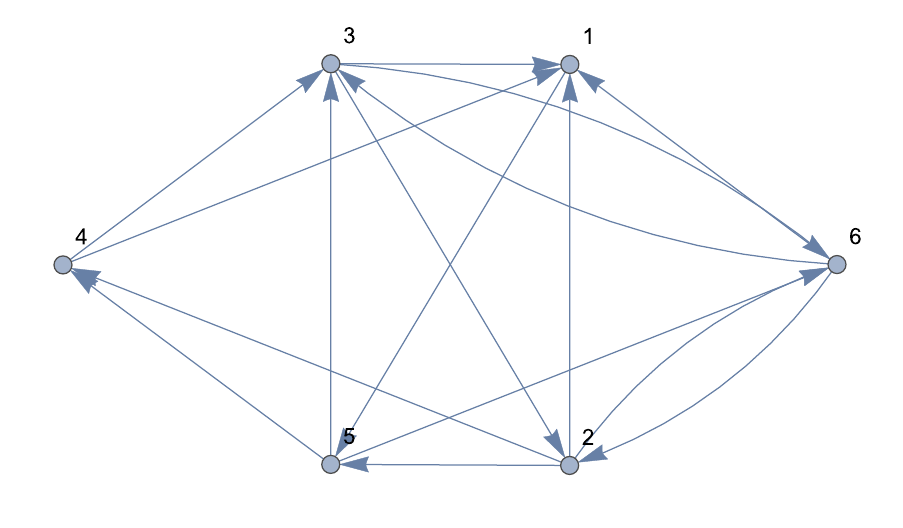}
\caption{Example of an Index Coding problem where the scheme in \cite[theorem 4]{arbabjolfaei2014local} are strictly larger than the proposed scheme for the fractional case. } 
\label{fig:comp_yhk}
\end{figure}

The proposed scheme of  \cref{thm:recur} is also better than the corresponding recursive scheme proposed in \cite[theorem 4]{arbabjolfaei2014local}. For the graph in \cref{fig:comp_yhk}
our scheme is strictly better than their scheme. The index coding broadcast rates for the graph are $3$ and $7/2$ for our proposed scheme and the scheme in \cite[theorem 4]{arbabjolfaei2014local}, respectively.

\vspace{0.1in}
\noindent{\em Remark (Codes with small alphabet size):}
Consider the construction of linear index codes using local chromatic number of the complement graph in \cite{shanmugam2013local}. Assume that the local chromatic number $\chi_l$ corresponds to a coloring with $m$ colors. Then, we require an $[m,\chi_l+1]$-MDS code. Using a Reed-Solomon code we can construct a $[m,\chi_l+1]$-MDS code on a field of size $m$. 
Here, we note that, instead of using the generator matrix of a $[m,\chi_l+1]$-MDS code the parity check matrix of any linear code of size $m$ and minimum distance $\chi_l+2$ would work. Thus, when restricted to using a small alphabet size (say $q$), we have the following upper-bound on the size of the code using the Gilbert-Varshamov bound,
$m-\log_q A_q(m,\chi_l+2) = m-\log_q \br*{\frac{q^m}{\sum_{j=0}^{\chi_l+1} \binom{m}{j}(q-1)^j}} = \log_q \br*{\sum_{j=0}^{\chi_l+1} \binom{m}{j}(q-1)^j}.$

\section{Further results: extension to k-GIC graphs}\label{sec:gic}

	\begin{figure}
	  \centering
	 \includegraphics[width=6cm]{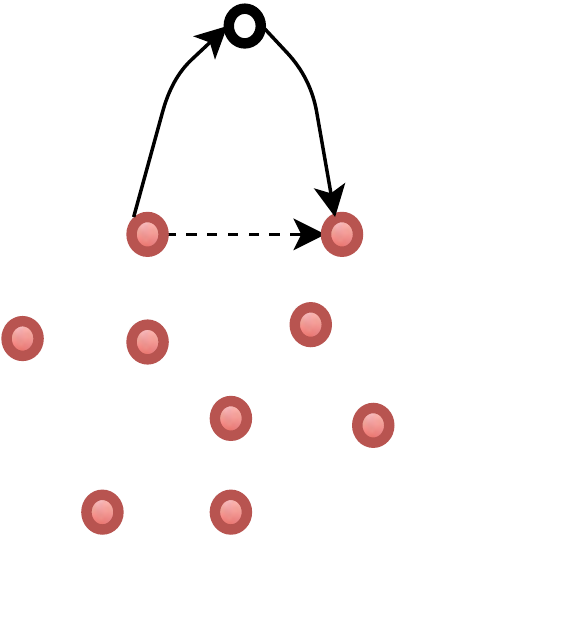}
	\caption{Example of $k$-GIC \cite{thapa2015generalized}. The vertices shown in red are the inner vertices. The inner vertex set contains all possible edges (not shown) within the subgraph except the dotted edge.}	
	\label{fig:GIC}
	\end{figure}

  In the previous section, we modified the scheme in \cite{shanmugam2013local} by substituting a $0$-partial clique cover of $\cG$ with a $k$-partial clique cover for any $k>0$.  We can further extend this procedure to include the $n$-$\mathsf{GIC}$ graphs defined in \cite{thapa2015generalized}. To that end, we extend the definition of $n$-$\mathsf{GIC}$ graphs to $(k,n)$-GIC, described below ($n$-$\mathsf{GIC}$ of \cite{thapa2015generalized} is $(0,n)$-GIC of our definition).


  Consider a directed graph with $ N $ vertices having the following properties:

1. A set of $ n $ vertices, denoted by $ V_{\mathrm{I}} $, such that for any vertex $v_i$ and at least $n-k-1$ vertices $v_j\in V_{\mathrm{I}}$ there is a path from $ v_i $ to $ v_j $ which does not include any other vertex of $ V_{\mathrm{I}} $. We call $ V_{\mathrm{I}} $ the \emph{inner vertex set}, and let $ V_{\mathrm{I}}=\{v_1,v_2,\dotsc,v_n\} $. The vertices of $ V_{\mathrm{I}}$ are referred to as \emph{inner vertices}. 

2. Due to the above property, we can always find a \emph{directed rooted tree}, (denoted by $ T_i $) with maximum number of leaves in $V_\rmI$ and root vertex $ v_i $, having at least $n-k-1$  other vertices in $V_\rmI \setminus \{v_i\} $ as leaves. The trees may be non-unique.

  Denote the union of all selected $ n $ trees as  $ D_n\triangleq  \bigcup_{\forall i\in {V_{\mathrm{I}}}} T_i $. If $D_n $ satisfies two conditions (to be defined shortly), we call it a $ (k,n) $-$ \mathsf{GIC} $ structure (denoted as a $ (k,n) $-$ \mathsf{GIC} $ sub-digraph: $ D_n=(V(D_n),E(D_n)) $, where  $ |V(D_n)|=N $). Now we define a type of cycle and a type of path.
  \begin{definition} [I-cycle]
  A cycle that includes only one inner vertex $ i\in V_{\mathrm{I}} $ is an \emph{I-cycle}. 
  \end{definition}
  \begin{definition} [P-path]
  A path in which only the first and the last vertices are from $ V_{\mathrm{I}} $, and they are distinct, is a \emph{P-path}. 
  \end{definition}
  
  The conditions for $ D_n $ to be qualified as a $(k,n)$-$\mathsf{GIC}$ are as follows:
  \begin{enumerate}
    \item \emph{Condition 1:} There is no \emph{I-cycle}.
    \item \emph{Condition 2:} For all ordered pairs of inner vertices ($ v_i,v_j $), $ i\neq j $, there is only one \emph{P-path} from $ i $ to $ j $. 
  \end{enumerate}

  \Cref{fig:GIC} shows the example of a $k$-GIC as defined in \cite{thapa2015generalized}. It is an ``almost" complete graph on the inner vertex set i.e. if all the P-paths are replaced with edges then the inner vertices form a $0$-partial clique. Our definition of $ (k,n) $-$ \mathsf{GIC} $ extends the definition in \cite{thapa2015generalized} such that when all the P-paths are replaced with edges, the inner vertex set $V_\rmI$ forms a $k$-partial clique.

  \subsection{Code Construction}
  The following theorem allows us to construct an index coding scheme. 
\begin{theorem} \label{thm:subtree} If a vertex $v \in V\setminus V_\rmI$ belongs to trees $T_i$ and $T_j$, $i\ne j$, then all the non-inner nodes on the subtree of $T_i$ rooted at $v$ also belong to $T_j$.
\end{theorem}
Note that, although \cref{thm:subtree} is similar to \cite[lemma 3]{thapa2015generalized} it is different in that it applies to $(n,k)$-GIC in constrast to \cite{thapa2015generalized} where applies only to $(n,0)$-GIC. The proof of this theorem is deferred to  \cref{appendix_a}. In particular in  \cref{appendix_a}, we show  that the proof in \cite[Lemma 3]{thapa2015generalized} still applies for the $(k,n)$-GIC. 

Denote the non-inner vertices of the subtree rooted at vertex $v \in V\setminus V_\rmI$ as $D_n(v)$. Let $[\bf{v}_1 \;\bf{v}_2\;\ldots \;\bf{v}_n]$ be the generator matrix of a $[n,k+1]$-MDS code. Then the index code is defined below:

1.  $\bf{w}_{\mathrm{I}} = \sum_{v_ji \in V_{\mathrm{I}}} \bf{v}_i x_i$ .

2.  Let $\bf{x}_{N(v_j,D_n)} \in \F_q^{\abs{N(v_j,D_n)}}$ denote the vector consisting of the data corresponding to the nodes in $N(v_j,D_n)$. A vector $\bf{w}_j \in \F_q^{\min\set{\abs{N(v_j,D_n)},k+1}}$ corresponding to all vertices $v_j \in V\setminus V_{\rm{I}}$ is transmitted, where,
    \begin{equation}
      \bf{w}_j = 
                \begin{cases} 
                     \sum_{v_l \in \set{N(v_j,D_n) \cap V_\rmI}} \bf{v}_{l} (x_j+ x_l)  +  \sum_{v_l \in \set{N(v_j,D_n) \setminus V_\rmI}}\\ \qquad \bf{u}_{l} (x_j + x_l) \hfill \text{if } \abs{N(v_j,D_n)}\geq k+1 \\
                     \bf{1} x_j + \bf{x}_{N(v_j,D_n)} \hfill  \text{if }\abs{N(v_j,D_n)}< k+1 \\
                \end{cases}
   \end{equation}
  where $\bf{u}_{l} \in \F_q^{\min\set{\abs{N(v_j,D_n)},k+1}}$ are chosen as described in \cref{algo_select_u}.
  \Cref{algo_select_u} first assigns a vector $\bf{u}_l$ to all non-inner nodes $v_l \in D_n$ starting from the nodes at the bottom of a tree $T_i \in D_n$. The vector $\bf{u}_l$ corresponding to the vertex $v_l$ depends on its children $\set{\bf{u}_l : v_l \in D_n(v_l)}$. To show that the algorithm works, we need to find a vertex for which all out-neighbors are outside $S$. It is easy to see from \cref{thm:subtree} that this always holds.


\begin{algorithm}
  \KwData{trees $T_1, T_2, \ldots, T_n$}
  \KwResult{$\bf{u}_{j} \in \F_q^{\min\set{\abs{N(v_j,D_n)},k+1}}$  for $v_j \in V\setminus V_\rmI$ and $\bf{u}_j \in \F^{k+1}$ for $v_j \in V_\rmI$}
  $\bf{u}_{i} = \bf{v}_i$ for all $v_i \in V_{\rmI}$\\
  $S =  V\setminus V_\rmI$\\
    \While{$\abs{S}>0$}{
      Find a vertex in $v_i\in S : N(v_i,D_n)\subseteq \overline{S}$ \\
      $\bf{u}_i = -\displaystyle\sum_{j:v_j\in N(v_i,D_n)}\bf{u}_{j}$\\
      $S = S\setminus \{v_i\}$
    }
  \caption{Selecting the vectors $\bf{u_j}, j\in V\setminus V_{\rm{I}}$}
  \label{algo_select_u}
\end{algorithm}

\subsection{Decoding}
It is easy to see that all the non-inner vertices $j\in V\setminus V_{\mathrm{I}}$ can recover their data $x_j$. We show that $v_i \in V_{\rm{I}}$ can also recover $x_i$.  Define ${\bf{w}_j}^\prime$ corresponding to the transmitted vector $\bf{w}_j$ for $v_j \in V\setminus V_\rmI$ as follows,
\begin{equation}
  {\bf{w}_j}^\prime = 
          \begin{cases} 
               \bf{w}_j \hfill &\text{if } \abs{N(v_j,D_n)}\geq k+1 \\
               [\bf{u}_{c_1}\; \bf{u}_{c_2}\; \ldots \;\bf{u}_{c_r}] \bf{w}_j \hfill & \text{if }\abs{N(v_j,D_n)}< k+1 \\
          \end{cases}
\end{equation}
where $\set{v_{c_1},v_{c_1},\ldots,v_{c_r}} = N(v_j,D_n)$. Denote by $T(v)$ the subtree rooted at vertex $v$ in tree $T$. For the non-inner child $v$ of vertex $v_i$ compute the following,
\begin{equation}
  \bf{w}(v) = \sum_{v_l \in T_i(v)\setminus V_{\rm{I}}} \bf{w}^\prime_l
\end{equation}
Note that, $\bf{w}(v)$ only contains terms of the form $\bf{u}_l x_l$ for $\set{l : v_l \in V_{\rm{I}} \text{ or } v_l \in D_n(v_i)}$. Now, consider
  $\bf{w}_\rmI - \sum_{v \in D_n(v_i)} \bf{w}(v)$.
The only terms left in this 
are $\bf{u}_l x_l$ for $\set{l : v_l \in V_{\rm{I}}\setminus T_i}$. Since there are at most $k$ such terms and $[\bf{v}_1 \;\bf{v}_2\;\ldots \;\bf{v}_n]$ is the generator matrix of a $[n,k+1]$-MDS code, vertex $v_i$ can decode $x_i$.

\section{Achievability scheme (proof of \cref{thm:main})}\label{sec:achiev_sch}




We now describe an index coding scheme to achieve a broadcast rate equal to the optimal solution of the  program in \cref{local_k_partial_integer_prog}.

Assume without loss of generality  that $\cS_1 = [n_1],\; \cS_2 = [n_1+1,n_1+n_2],\; \ldots, \cS_t = [\sum_{j\in[t-1]}n_j+1,\sum_{j\in[t]}n_j]$ be the partial cliques selected in \cref{local_k_partial_integer_prog}. Let $k_j \defeq k_{\cS_j}$. Assume that the optimum value of  \cref{local_k_partial_integer_prog} is $m$. Then $\max_j (k_j+1) \leq m \leq \sum_j (k_j+1)$. We construct $[n_j,k_j+1]$-MDS codes for each $j \in [t]$. Let $G_j$ denote the generator matrices for these codes. Thus, 
\begin{equation}\label{G_j}
  G_j = \begin{bmatrix}
    1 & 1 & \ldots & 1 \\
    \alpha_{j,1} & \alpha_{j,2} & \ldots & \alpha_{j,n_j} \\
    & & \vdots &\\
    \alpha_{j,1}^{{k_j}} & \alpha_{j,2}^{{k_j}} & \ldots & \alpha_{j,n_j}^{{k_j}} \\
  \end{bmatrix}
\end{equation}
where $\alpha_{j,1}, \alpha_{j,2}, \ldots, \alpha_{j,n_1}\in\F_q$ for all $j\in [t]$. 

Let $k^j = \sum_{l=1}^j (k_l+1)$. Assume that $\Phi$ is the generator matrix of an $[k^t,m]$-MDS code. Let $\Phi_j = \Phi_{\brac*{k^{j-1}+1,k^{j}}}$. Denote, $x^i_j = \min\set{\abs{\cS_j\cap \overline{N(v_i,\cG)}},(k_{\cS_j}+1)},\;\; i\in [n], j\in [t]$. 

Finally we construct the following matrix $G$,
\begin{equation}\label{code_construction}
	G = [\bf{u}_1\; \bf{u}_2\; \ldots \bf{u}_n] = [\Phi_1 G_1\; \Phi_2 G_2\; \ldots \Phi_t G_t]_{m \times n}
\end{equation}
where $n = \sum_{j} n_j$ is the number of vertices in $\cG$. 

Now assigning the vector $\bf{u}_i$ to vertex $i$, we show that this assignment satisfies the interference alignment condition in \cref{interference_align_condition}. Note that this scheme is a natural generalization of the localized coloring scheme where the matrix $G_j$ is the generator matrix of a repetition code and $k^t=m$.

{\lemma \label{full_rank} Let $H_j$ be a $(k_j+1)\times (k_j+1)$ submatrix  of $G_j$. Then for a large enough field $\F_q$ there exist $\alpha_{j,i}\in \F_q, j\in [t], i\in [n_j]$ such that $\hat{G}=[\Phi_1 H_1\;\Phi_2 H_2\;\ldots\;\Phi_t H_t]$ is an MDS matrix for any set of sub-matrices $H_j$.}
\begin{proof}
  Let $\Phi = [\bf{v}_1\;\bf{v}_2\;\ldots\;\bf{v}_{k^t}]$ and let $U(s)$ denote any $m\times s$ sub-matrix of $\Phi_{\geq2} \defeq [\Phi_2\;\ldots\;\Phi_t]$. Since $\Phi$ is MDS, $[\bf{v}_{i_1} \;\bf{v}_{i_2}\;\bf{v}_{i_r} \;U(m-r)]$ must be full-rank for all $\set{i_1, \ldots, i_r}\subseteq [k_1+1]$. Consider any vector $\bf{w}\in\F_q^m$,
  \begin{equation}\label{w_a}
    \bf{w} = [\bf{v}_{i_1} \;\bf{v}_{i_2}\;\bf{v}_{i_r} \;U(m-r)]\begin{bmatrix}
    \bf{a}_r\\
    \bf{a}_{r+1}
  \end{bmatrix}
  \end{equation}
  where $\bf{a}_r \in \F_q^r$ and $\bf{a}_{r+1} \in \F_q^{(m-r)}$. We show that there exist $\alpha_{1,i} \in \F_q, i\in [n_1]$ such that $\bf{w}$ can also be represented as a linear combination of column vectors in $G^\prime \defeq [\Phi_1 H_1 \;\;U(m-r)]$ where,
  \begin{equation*}
    H_1 = \begin{bmatrix}
      1 & 1 & \ldots &1  \\ 
      \alpha_{1,i_1} & \alpha_{1,i_2} & \ldots &\alpha_{1,i_r}  \\ 
      && \vdots \\
      {\alpha_{1,i_1}}^{k_1} & {\alpha_{1,i_2}}^{k_1} & \ldots &{\alpha_{1,i_r}}^{k_1}  \\ 
    \end{bmatrix}
  \end{equation*}
  for any $\set{i_1, i_2,\ldots,i_r} \subseteq [n_j]$, i.e.
  \begin{equation}\label{w_as_mds}
    \bf{w} = G^\prime\; \begin{bmatrix} \bf{d}_r \\ \bf{d}_{r+1} \end{bmatrix}
  \end{equation}
  for some $\bf{d}_r \in \F_q^r$ and $\bf{d}_{r+1}\in \F_q^{m-r}$.

  Since, $[\bf{v}_{i_1} \;\bf{v}_{i_2}\;\bf{v}_{i_r} \;U(m-r)]$ is full-rank, we have,
  \begin{equation}\label{v_r_1}
    [\bf{v}_{i^\prime_1}  \; \bf{v}_{i^\prime_2} \; \ldots\; \bf{v}_{i^\prime_{k_1+1-r}} ] =  [\bf{v}_{i_1} \;\bf{v}_{i_2}\;\bf{v}_{i_r} \;U(m-r)] \begin{bmatrix} {{B}_r} \\ {{B}_{r+1}} \end{bmatrix}, 
  \end{equation}
  where ${B}_r \in \F_q^{r\times\br{k_1+1-r}}$ and ${B}_{r+1} \in \F_q^{(m-r)\times\br{k_1+1-r}}$ and $\set{i^\prime_{1},i^\prime_{2},\ldots,i^\prime_{k_1+1-r}}=[k_1+1]\setminus\set{i_1,i_2,\ldots,i_r}$. Thus, combining \cref{w_a,w_as_mds,v_r_1}, we have,
  \begin{align}\label{final_cond}
      [\bf{v}_{i_1}\;&\bf{v}_{i_2}\;\ldots\;\bf{v}_{i_r}\; U(m-r)]\cdot \nonumber\\
     & \br*{\begin{bmatrix}
              [I_r\; B_r] \; P_\sigma\; H_1 \bf{d}_r \\
              B_{r+1} H_1^\prime \bf{d}_r + \bf{d}_{r+1}
            \end{bmatrix} - \begin{bmatrix}
              \bf{a}_{r}\\
              \bf{a}_{r+1}
            \end{bmatrix}} = \bf{0}
  \end{align}
  where,   
  \begin{equation*}
    H_1^\prime = \begin{bmatrix}
      \alpha_{1,i_1}^{r+1} & \alpha_{1,i_2}^{r+1} & \ldots &\alpha_{1,i_r}^{r+1}  \\ 
      \alpha_{1,i_1}^{r+2} & \alpha_{1,i_2}^{r+2} & \ldots &\alpha_{1,i_r}^{r+2}  \\ 
      && \vdots \\
      {\alpha_{1,i_1}}^{k_1} & {\alpha_{1,i_2}}^{k_1} & \ldots &{\alpha_{1,i_r}}^{k_1}  \\ 
    \end{bmatrix}
  \end{equation*}
  and $P_\sigma $ is the permutation matrix for $\sigma = (1\rightarrow i_1, 2\rightarrow i_2,\ldots, r\rightarrow i_r, (r+1)\rightarrow i_1^\prime, {r+2}\rightarrow i_2^\prime,\ldots, (k+1)\rightarrow i_{k+1-r}^\prime)$. For the solution in \cref{final_cond} to exist for all $\bf{a}_r \in \F_q^r, \bf{a}_{r+1} \in \F_q^{k-r+1}$ we must have,
$
    \det \br*{[I_r\; B_r] P_\sigma H_1} \ne 0
$
  i.e.,
  \begin{align} \label{field_size_condition}
  \det\left[
\begin{matrix}
     {\alpha_{1,i_1}}^{i_1-1}+\bf{g}({\alpha_{1,i_1}})\bf{b}_1 & {\alpha_{1,i_2}}^{i_1-1}+\bf{g}({\alpha_{1,i_2}})\bf{b}_1 &\ldots    &      {\alpha_{1,i_r}}^{i_1-1}+\bf{g}({\alpha_{1,i_r}})\bf{b}_1\\
     {\alpha_{1,i_1}}^{i_2-1}+\bf{g}({\alpha_{1,i_1}})\bf{b}_2 & {\alpha_{1,i_2}}^{i_2-1}+\bf{g}({\alpha_{1,i_2}})\bf{b}_2 &\ldots   &{\alpha_{1,i_r}}^{i_2-1}+\bf{g}({\alpha_{1,i_r}})\bf{b}_2\\
     &\vdots&\vdots & \vdots\\
     {\alpha_{1,i_1}}^{i_r-1}+\bf{g}({\alpha_{1,i_1}})\bf{b}_r & {\alpha_{1,i_2}}^{i_r-1}+\bf{g}({\alpha_{1,i_2}})\bf{b}_r &\ldots    & {\alpha_{1,i_r}}^{i_r-1}+\bf{g}({\alpha_{1,i_r}})\bf{b}_r\\
   \end{matrix} \right]
    \ne0
  \end{align}
  where $\bf{g}(\alpha)=[\alpha^{i^\prime_1-1} \alpha^{i^\prime_2-1} \ldots \alpha^{i^\prime_{k_1+1-r}-1}]$ and $B_r = \begin{bmatrix}
    \bf{b}_1\,\,
    \bf{b}_2 \,\,
    \hdots\,\,
    \bf{b}_r
  \end{bmatrix}^T.$

The polynomial in the RHS of \cref{field_size_condition} has degree at most $k_1$ for all the variables $\alpha_{1,i_1}, \alpha_{1,i_2},\ldots, \alpha_{1,i_r}$. Thus, by increasing the size of the field $\F_q$ we can make sure that there exist $\alpha_{j,i}$ for all $j\in [t]$ and $i \in [n_j]$ such that \cref{field_size_condition} holds. 

Now, repeating the above argument $t$ times we can say that $\left[\Phi_1 H_1 \Phi_2 H_2\;\Phi_3 H_3\;\ldots \;\Phi_t H_t\right] $ is MDS for all sets of $H_j \in \F_{q^r}^{(k_j+1)\times (k_j+1)}$ submatrices of $G_j$.

Whereas a loose upper-bound (sufficient) on the field size becomes, 
\begin{equation}\label{loose_upper_bound}
 q \leq \max_{j\in[t]} \sum_{r=1}^{k_j+1} k_j {{n-n_j}\choose {m-r}} {{n_j}\choose {r-1}} +n_j
\end{equation}
Now our main theorem is evident.
\end{proof}
{\theorem \label{interference_alignment_condition_scheme} For any vertex $v_i \in V(\cG)$ we have, 
\begin{equation}\label{eq1_thm}
  \bf{u}_i \not\in \spn\br{\bf{u}_{\overline{N(v_i,\cG)}\setminus{v_i}}}  
\end{equation}
}
\vspace{-0.2in}
\begin{proof}
  Consider a vertex $v_i$ in graph $\cG$, with non-neighbors $P_1, P_2, \ldots, P_t$ such that $P_j \subseteq \set{ i : v_i \in \cal{S}_j}$. Assume, wlog that $v_i\in \cS_1$. Then, $\abs{P_1}\leq k_1$. Note that, for any $P_j : \abs{P_j} \geq k_j+1$ there exists a set of $P^\prime_j \subseteq P_j : \abs{P^\prime_j} = k_j+1$ and $\spn\br*{\bf{u}_{P_j}} = \spn\br*{\bf{u}_{P^\prime_j}}$. Let $P^\prime_j = P_j$ for $P_j : \abs{P_j} \leq k_j+1, j\ne 1$ and $P^\prime_1 = P_1 \cup i$.  Therefore, we have 
  $\spn\br*{\set{\bf{u}_{P_j}}_j} = \spn\br*{\set{\bf{u}_{P^\prime_j}}_j}$.
  Now, since ${\set{\bf{u}_{P^\prime_j}}_j}$ is full-rank (lemma \ref{full_rank}) its easy to see that \cref{eq1_thm} follows.
\end{proof}

Note that, in the proof of \cref{interference_alignment_condition_scheme} we do not need the matrix $\hat{G}$ in \cref{full_rank} to be MDS. Instead we only need a class of $n$ different subsets of column vectors of $G$, each of size at most $m$, to be linearly independent. Thus the upper bound on the size of the alphabet in \cref{loose_upper_bound} is very loose and we can show that an alphabet of size $O(n)$ should suffice (details omitted).

The optimal index coding solution in \cref{local_k_partial_integer_prog} uses at most 2 levels of MDS codes $\Phi$ and $G_j, j\in [t]$. If we recursively use the above method on the $k$-partial cliques corresponding to $G_j$ then we can further reduce the index coding broadcast rate in \cref{local_k_partial_integer_prog}. The achievement scheme for the recursive linear program can be easily obtained by replacing $G_j$ in \cref{code_construction} with the matrix obtained for $IC_{LP}(\cG\rvert_\cS)$. The matrix for $IC_{LP}(\cG\rvert_\cS)$ can be obtained recursively in a similar manner. The integer program corresponding to the recursive scheme is,
\begin{equation}\label{local_partial_rec_integer_prog}
\begin{array}{ll@{}}
\text{minimize}  & t \\
\text{subject to}& \sum_{\cS \in \cK} \min\set{\abs{\cS\cap \overline{N(v,\cG)}},IC_{LP}(\cG\rvert_\cS)}\rho_\cS \le t, \\
 & \hspace{2in} v \in V(\cG)  \\ 
\text{and}& \displaystyle\sum_{\cS \in \cK : v \in \cS}  \rho_\cS \geq 1, \;\; v \in V(\cG)  \\
                 &                                          \rho_\cS \in \{0,1\}, \;\;\cS \in \cK
\end{array}
\end{equation}
where $IC_{LP}(\cG)$ denotes the optimum of the above integer program for the graph $\cG$ and $\cG\rvert_\cS$ denotes the graph $\cG$ restricted to the subset $\cS$. Note that $IC_{LP}(\cG\rvert_{\set{i}})\defeq 1, i\in [n] $ and the proof in lemma \ref{full_rank} and \cref{interference_alignment_condition_scheme} apply to the scheme for \cref{local_partial_rec_integer_prog} as well.

\section{Fractional local partial clique cover (proof of \cref{thm:frac})}\label{sec:frac}
In this section we show how to design an encoding scheme to achieve the rate given by the linear programming optimal solution of
\cref{local_k_partial_linear_prog}.

Since all the coefficients of the  LP in \cref{local_k_partial_linear_prog} are integers, the optimal solution $\rho_\cS^\star$ must be rational. Assume that in optimal solution to \cref{local_k_partial_linear_prog} all partial cliques $\cS$ for which $\rho_\cS>0$ are $\cS_1\; \cS_2,\; \ldots, \cS_t$ and $\abs{\cS_j}=n_j$. Let $k_j \defeq k_{\cS_j}$. Also assume that all $\rho_{\cS_j}, j\in [t]$ have a common denominator $\Delta$ such that $\rho_{S_j} = N_j/\Delta$. Assume that the optimum value to the optimization of \cref{local_k_partial_linear_prog} is $m^\star$. Construct a $[n_j,k_j+1]$-MDS code for each $j$. Let $G_j, j\in [t]$ be the generator matrix for these codes. Thus, 
\begin{equation}\label{G_j}
  G_{j} = \begin{bmatrix}
    1 & 1 & \ldots & 1 \\
    \alpha_{j,1} & \alpha_{j,2} & \ldots & \alpha_{j,n_j} \\
    & & \vdots &\\
    \alpha_{j,1}^{{k_j}} & \alpha_{j,2}^{{k_j}} & \ldots & \alpha_{j,n_j}^{{k_j}} \\
  \end{bmatrix}
\end{equation}
where $\alpha_{j,1}, \alpha_{j,2}, \ldots, \alpha_{j,n_1}\in\F$ for all $j\in [t]$. Let $G_j^\prime \defeq \sum_{l=1}^{N_j} Q_{ll}\otimes G_{j},$ where $Q_{rs}$ is a $N_j \times N_j$ matrix with the only non-zero entry being $Q_{rs}(r,s)=1$ and $\otimes$ denotes the matrix tensor product. Let $k^j = \sum_{l=1}^j N_j (k_l+1)$ and let $\Phi$ be the generator matrix of an $[k^t,m^\star \Delta]$-MDS code. Let $\Phi_j = \Phi_{\brac*{k^{j-1}+1,k^{j}}}$. Finally we construct a matrix $G$:
\begin{align}\label{code_construction_frac}
  G &= [\bf{u}_{1,1}\ldots\bf{u}_{1,(N_1n_j)}\; \cdots  \;\bf{u}_{t,1} \ldots \bf{u}_{t,N_tn_t}] \\
  &= [\Phi_1 G_1^\prime\; \Phi_2 G_2^\prime\; \ldots \Phi_t G_t^\prime]_{(\Delta m^\star) \times (\Delta n)}
\end{align}
Note that $\sum_{j=1}^t n_j N_j = \Delta n$. Now each vertex has to be assigned $\Delta$ column vectors from $G$. Assume wlog that a vertex belongs to partial cliques $S_{1}, S_{2}, \ldots, S_{r}$, $r\leq t$. Assume that the column vector in $G_{j}, j\in [r]$ corresponding to the vertex is $\begin{bmatrix}
    1 \,\,
    \alpha_{j,c_j} \,\,
    \hdots\,\,
    \alpha_{j,c_j}^{k_{j}}
  \end{bmatrix}^T$, $c_{j} \in [n_{j}]$. Then the vectors assigned to the vertex are $\bf{u}_{j, c_j}, \bf{u}_{j, c_j+n_j}, \bf{u}_{j,c_j+2n_j},\ldots, \bf{u}_{j,c_j+(N_j-1)n_j}$ for $ j\in [r]$.

Now using  arguments similar to \cref{full_rank}, the interference alignment condition can be satisfied for this case too.

To construct an achievability for the fractional relaxation of the recursive scheme (\cref{local_partial_rec_lin_prog}), we can recursively use the scheme for \cref{local_k_partial_linear_prog}.
\begin{equation}\label{local_partial_rec_lin_prog}
\begin{array}{ll@{}}
\text{minimize}  & t \\
\text{subject to}& \sum_{\cS \in \cK} \min\set{\abs{\cS\cap \overline{N(v,\cG)}},IC_{FLP}(\cG\rvert_\cS)}\rho_\cS \le t, \\
 & \hspace{2in} v \in V(\cG)  \\ 
\text{and}& \displaystyle\sum_{\cS \in \cK : v \in \cS}  \rho_\cS \geq 1, \;\; v \in V(\cG)  \\
                 &                                          \rho_\cS \in [0,1], \;\;\cS \in \cK
\end{array}
\end{equation}
where $IC_{FLP}(\cG)$ denotes the optimum value  of the above linear program for graph $\cG$. Let a vertex $i \in [n]$ belong to $r$ recursive index coding problems $IC_{LP}(\cG), IC_{LP}(\cG\rvert_{\cS_1}),\ldots, IC_{LP}(\cG\rvert_{\cS_r})$, $\cG \supset \cS_1 \supset \cS_2 \ldots \supset \cS_r$. Thus, for the achievability scheme above, it is assigned $r$ coefficients corresponding to the $r$ problems, $\rho_1 = N^1/\Delta^1, \ldots, \rho_r = N^r/\Delta^r$ and assigned $\Delta^1 \times \ldots \times \Delta^r$ column vectors. 

Note that, we need the size of the node, $\ell$, to be at least $\Delta^1 \times \ldots \times \Delta^r$, for the scheme to work.

\bibliography{../references}

\appendix
\crefalias{section}{appsec}
\subsection{A Simpler Achievability Scheme for the Recursive Scheme in \cite[theorem 4]{arbabjolfaei2014local}}\label[Appendix]{appendix_b}

Consider the following integer program proposed in \cite{arbabjolfaei2014local},
\begin{equation}\label{local_partial_integer_prog_yhk}
\begin{array}{ll@{}}
\text{minimize}  & t \\
\text{subject to}& \sum_{\cS \in \cK} (k_\cS+1) \rho_\cS \le t, \\
 & \hspace{2in} v \in V(\cG)  \\ 
\text{and}& \displaystyle\sum_{\cS \in \cK : v \in \cS}  \rho_\cS \geq 1, \;\; v \in V(\cG)  \\
                 &                                          \rho_\cS \in [0,1], \;\;\cS \in \cK.
\end{array}
\end{equation}

We propose a simple scheme corresponding to the integer program in \cref{local_partial_integer_prog_yhk}. Assume that the optima corresponding to the integer program is $m$, the partial cliques selected are $\cS_1, \cS_2, \ldots, \cS_t$. Let $n_j \defeq \abs{\cS_j}$  and $k_j \defeq k_{\cS_j}$. Let the cliques selected corresponding to the vertex $v_i$ have indices $\set{c_1^i, c_2^i, \ldots, c_{t_i}^i} \subseteq [t]$.

We construct the following matrix $G$,
\begin{equation}
  G = [\bf{u}_1\; \bf{u}_2\; \ldots \bf{u}_n] = [\Phi_1 G_1\; \Phi_2 G_2\; \ldots \Phi_t G_t]_{m \times n}
\end{equation}
where $n = \sum_{j} n_j$ is the number of vertices in $\cG$, $G_j, j\in [t]$ are $[n_j,k_j+1]$-MDS matrices, and $\Phi_j$ are $m\times (k_j+1)$ matrices defined as follows.

Let $\Phi$ denote a set of $m$ linearly independent vectors of dimension $m$. We construct the matrix $\Phi_j$ such that its column vectors are vectors in $\Phi$ and the column vectors for the matrices $\Phi_{c^i_1}, \ldots, \Phi_{c^i_{t_i}}$ are full rank for all $i$. Note that such an assignment is possible because  $m = \max_i \sum_{j=1}^{t_i} (k_{c_j^i}+1)$.

Note that, in this case, the upper bound on the alphabet size is $\max_{j\in [t]} n_j$ corresponding to the largest alphabet size needed for constructing $G_j$.

The achievability scheme for the recursive scheme in \cite[theorem 4]{arbabjolfaei2014local} can be constructed from \cref{local_partial_integer_prog_yhk} analogous to the way for which we construct the scheme for \cref{local_partial_rec_integer_prog} from \cref{local_k_partial_integer_prog}. And the extension to linear programs is also done in an analogous manner.

\subsection{Proof of \cref{thm:subtree}}\label[Appendix]{appendix_a}

Denote the leaves of subtree of the tree $T$ rooted at vertex $v \in V(T)$ as $L(v,T)$ and the leaves of the tree $T$ as $L(T)$.
{\lemma \label{subtree_lemma_prelim} If a vertex $v\in V(T_i) \text{ and } v\in V(T_j), \;i\ne j$ and $v\in V_\rmI$ then  $L(v,T_j), L(v,T_i)\subseteq V_\rmI \setminus \set{v_i,v_j}$}

\begin{proof}
  If the vertex $ v_j\in L (v,T_i) $, then there exists a path from vertex $ v $ to $ v_j $ in the tree $ T_i $. However, in the tree $ T_j $, there is a path from vertex $ v_j $ to $ v $. Thus in the sub-digraph $ D_n $, we obtain a path from vertex $ v $ to $ v_j $ (via $ T_i $) and vice versa (via $ T_j $). As a result, an $ I $-cycle containing $ v_j $ is present. This contradicts the condition 1 (i.e., no $ I $-cycle) for a $ D_n $. Hence, $ v_j\notin L (v,T_i) $. In other words, $ L (v,T_i) \subseteq V_\rmI \setminus \{v_i,v_j\} $. Similarly, $ L (v,T_j) \subseteq V_\rmI \setminus \{v_i,v_j\} $. 
\end{proof}

{\lemma \label{subtree_lemma1}If a vertex $v\in V(T_i) \text{ and } v\in V(T_j), \;i\ne j$ and $v\in V_\rmI$ then  $L(v,T_i)=L(v,T_j)$}
\begin{proof}
  From \cref{subtree_lemma_prelim}, $ L(v,T_i) $ is a subset of $ V_\rmI \setminus \{v_i,v_j\} $. Now pick a vertex $ v_c $ belongs to $ V_\rmI \setminus \{v_i,v_j\} $ such that $ v_c \in L(v,T_i)$ but $ v_c\notin L(v,T_j) $ (such $ v_c $ exists since we suppose that $ L(v,T_i)\neq L(v,T_{j}) $). In tree $ T_i $, there exists a directed path from the vertex $ v_i $ which includes the vertex $ v $, and ends at the leaf vertex $ v_c $. Let this path be $ P_{{v_i}\rightarrow v_c}(T_i)$.

  Now, suppose that in tree $ T_j $, there exists a directed path from the vertex $ v_j $, which doesn't include the vertex $ v $ (since $ v_c\notin L(v,T_j) $), and ends at the leaf vertex $ v_c $. Let this path be $ P_{{v_j}\rightarrow v_c}(T_j) $. However, in the digraph $ D_n $, we can also obtain a directed path from the vertex $ v_j $ which passes through the vertex $ v $ (via $ T_j$), and ends at the leaf vertex $ v_c $ (via $ T_i $). Let this path be $ P_{{v_j}\rightarrow c}(D_n) $. The paths $ P_{{v_j}\rightarrow v_c}(T_j) $ and $ P_{{v_j}\rightarrow v_c}(D_K)$ are different which indicates the existence of multiple $ P $-paths from the vertex $ v_j $ to $ v_c $ in $ D_n $, this contradict the condition 2 for a $ D_n $. Consequently, $ L(v,T_i)=L(v,T_j)$.

  Therefore, the only case left is $v_c \not\in L(T_j)$. But since, the tree $T_j$ must be such that it has the maximum number of leaves in $V_\rmI$ and there exists a tree that has more leaves than $T_j$ this leads to a contradiction.
\end{proof}

{\lemma \label{subtree_lemma2}If a vertex $v\in V(T_i) \text{ and } v\in V(T_j), \;i\ne j$ and $v\in V_\rmI$ then the out-neighborhood of the vertex $v$ must be same in both the trees i.e. $N(v,T_i)=N(v,T_j)$}
\begin{proof}
  Now we pick a vertex $ v_b $ such that, without loss of generality, $v_b\in N (v,T_i) $ but $v_b\not\in N(v,T_j) $ (such $ v_b $ exists since we assumed that $ N(v,T_i)\neq N(v,T_j) $). Furthermore, we have two cases for $ v_b $, which are (case 1) $ v_b\in L(v,T_i)$, and (case 2) $ v_b\notin L(v,T_i)$. Case 1 is addressed in \cref{subtree_lemma1}. On the other hand, for case 2, we pick a leaf vertex $ v_d \in L(v_b,T_i)$ such that there exists a path that starts from $ v $ followed by $ v_b $, and ends at $ v_d $, i.e., $ \langle v,v_b,\dotsc,v_d \rangle $ exists in $ T_i $. A path $ \langle v_j,\dotsc,v \rangle $ exists in $ T_j $. Thus a path $ \langle v_j,\dotsc,v,v_b,\dotsc,v_d \rangle $ exists in $ D_n $. From the first part of the proof, we have $ L(v,T_i)=L(v,T_j)$, so $ v_d\in L(v,T_j) $. Now in $ T_j $, there exists a path from $ v_j $ to $ v_d $, which includes vertex $ v $ followed by a vertex $ v_e$ such that $ v_e\in N (v,T_j)$ and $v_e\neq v_b$ (as $ v_b\notin N (v,T_j) $), and the path ends at $ v_d $, i.e., $ \langle v_j,\dotsc,v,v_e,\dotsc,v_d \rangle $ which is different from $ \langle v_j,\dotsc,v,v_b,\dotsc,v_d \rangle $. So multiple $ P $-paths are observed at $ v_d $ from $ v_j $. This contradicts condition 2 for a $ D_n $. Consequently, $ N(v,T_i)= N(v,T_j) $.  
\end{proof}

\end{document}